\documentclass[aps,prc,twocolumn,floatfix,showpacs,nofootinbib]{revtex4-1}

\usepackage{graphicx,amsfonts,amssymb,amsbsy}
\usepackage{amsmath,amsthm,latexsym}
\usepackage{ulem}

\usepackage{natbib}
\usepackage[colorlinks]{hyperref}
\hypersetup{linkcolor=blue,citecolor=blue}

\begin{document}

\title{Discontinuity gravity modes in hybrid stars: \\ 
assessing the role of rapid and slow phase conversions}

\author{L. Tonetto$^{1}$ and G. Lugones$^{1}$}

\affiliation{$^1$ Universidade Federal do ABC, Centro de Ci\^encias Naturais e Humanas, Avenida dos Estados 5001- Bang\'u, CEP 09210-580, Santo Andr\'e, SP, Brazil.}

\begin{abstract}
Discontinuity gravity modes may arise in perturbed quark-hadron hybrid stars when a sharp density jump exists in the stellar interior and are a potential fingerprint to infer the existence of quark matter cores in compact objects. When a hybrid star is perturbed, conversion reactions may occur at the quark-hadron interface and may have a key role in global stellar properties such as the dynamic stability and the quasi-normal mode spectrum. In this work we study the role of the conversion rate at the interface.  To this end, we first derive the junction conditions that hold at the sharp interface of a non-radially perturbed hybrid star in the case of slow and rapid conversions. Then, we analyse the discontinuity $g$-mode in both cases. For rapid conversions, the discontinuity $g$-mode has zero frequency because a displaced fluid element near the phase splitting surface adjusts almost immediately its composition to its surroundings and gravity cannot provide a buoyancy force.  For slow conversions, a $g$-mode exists and its properties are analysed here using modern hadronic and quark equations of state.  Moreover, it has been shown recently that in the case of slow conversions an extended branch of stable hybrid configurations arises for which $\partial M/ \partial \epsilon_c <0$.  We show that $g$-modes of the standard branch (that is, the one with $\partial M/ \partial \epsilon_c > 0$) have frequencies and damping times in  agreement with previous results in the literature. However,  $g$-modes of the extended branch have significantly larger frequencies (in the range $1-2 \, \mathrm{kHz}$) and much shorter damping times (few seconds in some cases).  We discuss the detectability of $g$-mode GWs with present and planned GW observatories. 
\end{abstract}

\maketitle

\section{Introduction} \label{sec:intro}

Since their discovery more than 50 years ago, neutron stars (NS) have attracted much attention because of their extreme physical properties. Such interest has been highly boosted recently by the direct detection of gravitational waves (GWs) from the NS merger GW170817 \citep{abbott2017b} and its electromagnetic counterparts GRB170817A and AT2017gfo \citep{abbott2017c} which imposed a new set of observational constrains on some key properties of these objects \citep{Raithel,Annala,Malik,Most,Fattoyev,Paschalidis:2017qmb,Tews:2018iwm,Christian:2019qer}. 

It is known that NSs contain matter under extreme conditions, but the exact nature of their deep interiors is still one of the key unsolved issues in the area.  It is therefore very important to identify astrophysical signatures that  can be unequivocally associated with specific internal aspect of NSs. For instance,  if a sharp density discontinuity  is present inside a NSs,  it has been proposed that the so-called gravity pulsation mode (hereafter $g$-mode) would arise when the star is perturbed (see Refs. \cite{finn1986,finn1987,finn1988}).  This is very relevant to understand the interiors of the so-called  hybrid stars, composed by a quark matter core and external hadronic layers.  In fact, since the $g$-mode of perturbed quark-hadron hybrid stars is expected to emit GWs in the ballpark of  0.5 kHz \cite{sotani2001,miniutti2003,floreslugones2014,Orsaria_2019}  this could work as a smoking gun to infer the presence of quark matter inside some nearby compact objects. 

In this work, we will investigate $g$-modes of quark-hadron hybrid stars assuming that the quark matter core is separated from the external hadronic layers of the star by a sharp interface with a density jump across which thermal, mechanical and chemical equilibrium is maintained.
The assumption of a sharp interface is expected to be correct  if charge screening and surface effects are high enough and inhibit the formation of a mixed phase of quarks and hadrons (see \citep{Voskresensky,Endo,Yasutake,Lugones_Grunfeld_Ajmi_2013,Lugones_Grunfeld_2017} and references therein).

A key aspect of a sharp interface that deserves a detailed investigation is its behaviour under small perturbations. When a hybrid star is perturbed, fluid elements all along the stellar interior are displaced from their equilibrium positions. In particular, fluid elements in the neighbourhood of the quark–hadron interface can be periodically compressed and rarified and their pressures may become  higher or lower than the phase transition pressure,  making possible a phase conversion.  However,  the quark-hadron conversion in a compact star involves a quite complex mechanism where strong interactions, surface and curvature effects, Coulomb screening, etc. play a significant role (see \cite{Lugones_Grunfeld_2017} and references therein); thus, a fluid element oscillating around the interface will not necessarily undergo a phase conversion. In fact, the probability that such conversion occurs depends on the nucleation timescale, which at present is a model dependent quantity with an uncertain value \cite{Lugones_Grunfeld_2017,Bombaci2016,lugones2016}.  
As shown in Fig. 11 of \cite{Bombaci2016}, typical timescales for nucleation driven by quantum fluctuations are much larger than the age of the Universe for temperatures below 10 MeV. For thermal nucleation the timescale is significantly larger than the age of the Universe for $T< 5$ MeV but decreases to $\sim 1 \,\mathrm{s}$ for $T \approx 10$ MeV. These numbers are significantly larger that the typical period of fluid oscillation modes in compact stars  (roughly $10^{-3} \mathrm{s}$), strongly suggesting that fluid elements oscillating around the quark-hadron interface will not undergo any phase conversion in a \textit{cold} compact star. However, due to the uncertainties involved in these model dependent calculations, our analysis will not exclude the possibility of faster nucleation timescales. Moreover, although results for temperatures above 11 MeV are not shown in \cite{Bombaci2016}, a rough extrapolation of their results for higher temperatures suggests that the nucleation time could be of the order of the oscillation period at temperatures about $\sim 20 \, \mathrm{MeV}$ or even below, suggesting that rapid conversions could be of interest in hot objects.

%

In recent works \cite{pereirafloreslugones2018,Mariani:2019vve} we analysed the stability of hybrid stars under small radial perturbations,  focusing on two limiting cases: \textit{slow} and \textit{rapid} phase conversions at the sharp interface. Slow conversions  involve the stretch and squash of volume elements near the quark-hadron interface without their change of nature (nucleation timescales much larger than those of perturbations). Rapid conversions imply a practically immediate conversion of volume elements from one phase to the other and vice-versa in the vicinity of the discontinuity upon any perturbation \cite{Haensel1989,pereirafloreslugones2018}. 
One of the main conclusions emerging from the analysis  is that the usual static stability condition  $\partial M / \partial \epsilon_c \ge 0$, where $\epsilon_c$ is the central density of a star whose total mass is $M$, always remains true if phase conversions are rapid but \textit{breaks down} in general if they are slow. As a consequence, an additional branch of stable HS configurations is possible in the case of slow phase conversions \cite{pereirafloreslugones2018,Mariani:2019vve}.

The main goal of the present work is to study the role of  \textit{slow} and \textit{rapid} phase conversions on discontinuity $g$-modes of zero-temperature hybrid stars.
The paper is organised as follows. In Section \ref{sec:TOV} we review the role of slow and rapid phase conversions on the stability of static configurations under small radial perturbations, and present the non-radial oscillation equations that will be employed in this work.  In Section \ref{sec:junction} we derive the junction conditions that hold at the sharp interface of a hybrid star when non-radial oscillations occur. In Section \ref{sec:none} we analyse the physical mechanism that suppresses the existence of discontinuity $g$-modes when phase conversions at the interface are rapid. In Section \ref{sec:results} we calculate the properties of $g$-modes in the case of slow phase conversions using specific equations of state (EOS) for hadronic and quark matter. Special attention in given to $g$-modes of objects located at the extended branch of hybrid stars.  Finally, in Section \ref{sec:conclusion} we summarise our results and explore some of their astrophysical consequences.

\section{Basic equations} \label{sec:TOV}

In order to calculate the oscillation modes of a compact star, its equilibrium structure has to be determined first using the stellar structure equations of Tolman-Oppenheimer-Volkoff (TOV)  \cite{TOV}. These  equations must be supplemented with an EOS which for cold catalyzed matter has the form $p=p(\epsilon)$, where $p$ is the pressure and $\epsilon$ the energy density. Since we focus on quark-hadron hybrid stars with a sharp density discontinuity in thermal, chemical and mechanical equilibrium, the EOS has the generic form:
\begin{equation}
\epsilon(p)= \begin{cases}  \epsilon_H(p)  &   p < p_t \\
                        \epsilon_Q(p)  &   p > p_t
\end{cases}
\end{equation}
where the subscript $H$ refers to hadrons, the subscript $Q$ to quarks, and  $p_t$ is the pressure at the density discontinuity (transition pressure). 

In Sec. \ref{subsec:stability} we discuss the role of the reaction speed at the sharp discontinuity on the stability of equilibrium configurations of compact stars. In Sec. \ref{sec:NRO} we present the equations for non-radial oscillations used in this work.

\subsection{Conversion speed at the interface and the dynamical stability of hybrid stars}
\label{subsec:stability}

The stability of an equilibrium stellar configuration can be analysed by its response to small radial perturbations \citep{chandrasekhar1964}. When a \textit{stable} configuration is perturbed, fluid elements all along the stellar interior oscillate around their equilibrium positions, compressing and expanding periodically. On the contrary, in the case of an \textit{unstable} configuration, small perturbations grow without limit, leading to the collapse or disruption of the star.
In the case of hybrid stars, special care must be taken with fluid elements close to the quark-hadron interface,  because, as the fluid oscillates, their pressures become alternatively higher and lower than the phase transition pressure $p_t$.  Such compression and decompression around the interface leads to two essentially different behaviours depending on the speed of the quark-hadron nucleation mechanism (which depends on many poorly known microphysical details). 	
If the nucleation timescale is much shorter than those of perturbations (rapid conversions) fluid elements will convert almost immediately from one phase to the other as their pressure go  alternatively beyond and below $p_t$. On the contrary, if the nucleation  timescale is much larger than those of perturbations (slow conversions) the motion around the interface involves only the stretch and squash of volume elements without any phase transition.

In  a recent work \citep{pereirafloreslugones2018}, we have shown that in spite of being physically complex, the nature of the conversion can be mathematically summarised into simple \textit{junction conditions} on the radial fluid displacement $\xi$ and  the corresponding Lagrangian perturbation of the pressure $\Delta p$ at the phase-splitting surface. 

For slow conversions, the jump of $\xi$ and $\Delta p$ across the interface should always be null:
\begin{equation}
[\xi ]^+_{-}\equiv \xi^+-\xi^- =0, \quad [\Delta p]^+_- \equiv \Delta p^+ - \Delta p^- =0.
\label{xislow}
\end{equation}
In the latter equation and in the rest of this work,  $[x]^{+}_{-} \equiv x^+ - x^-$ indicates the jump of quantity $x$ across the interface, being $x^+$ the value of $x$ just above the interface  (i.e. at the hadronic side) and  $x^-$ the value just below it (i.e. at the quark side).

For rapid phase transitions it was found that
\begin{equation}
[\xi^r]^+_-=\Delta p\left [\frac{1}{p'_0}\right]^+_- ,  \qquad [\Delta p]^+_- =0  ,
\label{xirapid}
\end{equation}
where $p_0' \equiv dp_0/dr$ is the pressure gradient of the background pressure at the interface. 

The formalism of small radial perturbations of spherically symmetric stars \citep{chandrasekhar1964} shows that, if the fundamental oscillation frequency is a real number ($\omega_{\mathrm{radial},0}^{2}>0$), then any radial perturbation of the star will produce oscillatory fluid displacements and the stellar configuration is dynamically stable.

However, in many cases, an equivalent and much simpler static condition can be derived from the latter one. In fact, it is widely known that for a hydrostatic configuration of a spherically symmetric cold-catalysed one-phase star it holds
\begin{equation}
\partial M/\partial \epsilon_c < 0   \quad \implies \quad \text{unstable configuration},
\label{dMdE_c}
\end{equation}
where $\epsilon_c$ is the central density of a star whose total mass is $M$ \citep{HTWW1965}. 
This result is closely related to the fact that an equilibrium configuration of cold catalysed matter possesses a characteristic mode of  vibration of zero frequency $(\omega_{\mathrm{radial},0}^2=0)$ when and only when $\partial M/ \partial \epsilon_{c}=0$; that is, changes of stability occur only at  critical points in the $M$ versus $\epsilon_{c}$ curve. In fact, the situation of having an equilibrium configuration for $A$ baryons near another equilibrium configuration  for the same number of baryons is ordinarily impossible for cold catalysed matter. Only at a critical point in the $M$ versus $\epsilon_{c}$ plane does there exist a displacement which carries the system from an equilibrium configuration to a nearby equilibrium of the same baryon number, which is the necessary and sufficient condition for a mode of zero frequency. 

Eq. \eqref{dMdE_c} is widely used in the literature, but it must be emphasised that its validity cannot be extended to situations where the hypothesis of the theorem are not fulfilled, e.g. when matter is non-catalysed.  
%
This can be simply understood as follows:  if fluid elements of a star with $M=M_{\max}$ undergo small radial perturbations and they are not able to attain chemical equilibrium, then the perturbed state cannot be located along the $M$ versus $\epsilon_{c}$ curve as it would be for cold catalysed matter. This means that, a star with $M_{\max}$ cannot be displaced horizontally in the $M$ versus $\epsilon_{c}$ plane if matter is non-catalysed and consequently the perturbed state cannot have the same gravitational mass as the unperturbed one, i.e. the star will not be able to remain indefinitely at the perturbed state. Therefore, the vibration of zero frequency $(\omega_{\mathrm{radial},0}^{2}=0)$ will not occur when $\partial M/ \partial \epsilon_{c}=0$ if matter is non-catalysed and the sign of $\partial M/ \partial \epsilon_{c}=0$ cannot be used to assess stellar stability. 
%
Some examples of this situation are already known in the literature. Gourgoulhon et al. \citep{Gourgoulhon} studied perturbations of NSs where the composition of matter is considered frozen because the time after which the composition of the perturbed state reaches its equilibrium nuclear composition is larger than the dynamical timescale of perturbations. Their linear analysis showed that stable NSs could exist with the central density higher than that corresponding to $M_{\max}$. 

Systems containing multiple phases separated by sharp density discontinuities \cite{pereirafloreslugones2018,Pereira2,Mariani:2019vve} are another case where  Eq. \eqref{dMdE_c}  may be invalid. In this context, we have shown in \cite{pereirafloreslugones2018,Mariani:2019vve},  that the standard stability criterion of Eq.~\eqref{dMdE_c} remains always true for rapid phase transitions {(which assume chemical equilibrium at all times in view of the very fast conversion rates)} but breaks down in general for slow phase transitions {(the volume elements at the phase-splitting interface do not reach chemical equilibrium when perturbed, they remain with the same composition)}. In fact, for slow transitions the frequency of the fundamental mode can be a real number (indicating stability) even along the branch of stellar models that verifies $\partial M/\partial \epsilon_c < 0$.  Thus, in the case of slow conversions, branches that were believed to be radially unstable are in fact radially stable under small perturbations.

It is worth emphasising that these results have been confirmed independently by \cite{DiClemente2020} using a different numerical method than the one employed in \cite{pereirafloreslugones2018,Mariani:2019vve}.  They  find that with any speed of sound and/or matter density discontinuity the last stable configuration is realized at a central energy density exceeding that of the maximum mass configuration. Their calculations confirm the results of \cite{pereirafloreslugones2018,Mariani:2019vve} for hybrid stars and extend them to any piecewise polytropic solutions, even in the presence of only a speed of sound discontinuity. 

%
%
%
%
%
%

\subsection{Non-radial oscillations} 
\label{sec:NRO}

In this work we are interested in the influence of some microphysical properties of matter on the GWs emitted by a compact object; therefore, we consider only even-parity perturbations, which are coupled to the fluid \citep{thornecampolattaro1967,thornecampolattaro1967erratum}.  The perturbed metric can be written as \citep{lindblomdetweiler1985}
\begin{equation}
\begin{aligned}
 ds^{2}= &-e^{2\nu} (1+r^l H_0 Y^l_m e^{i\omega t}) dt^{2}  \\
         & - 2i\omega r^{l+1} H_1 Y^l_m e^{i\omega t} dtdr \\
         & + e^{2\lambda}(1-r^l H_0 Y^l_m e^{i\omega t}) dr^2  \\
         & + r^{2}(1-r^l K Y^l_m e^{i\omega t})(d\theta^{2}+\sin^{2}{\theta}d\phi^{2}),
\label{eq:metric_DL}
\end{aligned}
\end{equation} 
where $Y_{lm}(\theta, \phi)$ are the spherical harmonic functions. The small amplitude motion of the perturbed configuration is described by the Lagrangian 3-vector fluid displacement $\xi^j$, which can be represented in terms of perturbation functions $W(r)$ and $V(r)$ as
\begin{eqnarray}
 \xi^r&=&r^{l-1} e^{-\lambda} W(r) Y^l_m e^{i\omega t}, \label{eq:Xi} \\
 \xi^\theta&=&-r^{l-2} V(r) \partial_\theta Y^l_m e^{i\omega t}, \\
 \xi^\phi&=&-r^l (r \ \sin\theta)^{-2} V(r) \partial_\phi Y^l_m e^{i\omega t}.
\label{eq:lagrangian_3vector}
\end{eqnarray}
Our analysis will be restricted to the $l = 2$ component, which dominates the emission of GWs. Introducing the variable $X$, defined by:
\begin{equation}
    X = \omega^2 (\epsilon+p) e^{-\nu} V - r^{-1} p,_r e^{(\nu-\lambda)} W + \frac{1}{2} (\epsilon+p) e^{\nu} H_0,
    \label{eq:definitionX}
\end{equation}
where $,_r$ indicates differentiation with respect to $r$, one can write a
fourth-order system of linear equations for $\left(H_{1}, K, W, X\right)$ \citep{lindblomdetweiler1983,lindblomdetweiler1985}:
\begin{eqnarray}
\label{eq:osc_system}
K,_r &=& H_0/r + \tfrac{1}{2}l(l+1)r^{-1}H_1 - [(l+1)/r-\nu,_r]K  \nonumber \\
&& - 8 \pi (\epsilon+p) e^{\lambda} r^{-1} W, \label{eq:oscK} 
\end{eqnarray}
\begin{eqnarray}
H_1,_r &=& -r^{-1} [l+1+2M e^{2\lambda}/r + 4 \pi r^2 e^{2\lambda} (p-\epsilon)] H_1 \nonumber \\
&& + r^{-1} e^{2 \lambda} [H_0 + K - 16 \pi (\epsilon+p) V], \label{eq:oscH1} 
\end{eqnarray}
\begin{eqnarray}
W,_r &=& -(l+1)r^{-1}W + r e^{\lambda} [(\gamma p)^{-1} e^{-\nu} X \nonumber \\
&& - l(l+1)r^{-2} V + \tfrac{1}{2} H_0 + K], \label{eq:oscW}
\end{eqnarray}
\begin{eqnarray}
X,_r &= & -l r^{-1} X + (\epsilon+p)e^{\nu} \{ \tfrac{1}{2} (r^{-1}-\nu,_r) H_0 \nonumber \\
&& + \tfrac{1}{2} [r \omega^2 e^{-2\nu}  + \tfrac{1}{2} l(l+1)/r] H_1 + \tfrac{1}{2} (3 \nu,_r - r^{-1})K  \nonumber \\
&&- l(l+1) \nu,_r r^{-2} V - r^{-1} [4 \pi (\epsilon+p) e^{\lambda}   \nonumber \\
&& + \omega^2 e^{\lambda-2\nu}- r^2 (r^{-2}e^{-\lambda} \nu,_r),_r] W \} \label{eq:oscX}.
\end{eqnarray}
In the above equations, $\gamma$ is the adiabatic index defined by
\begin{equation}
    \gamma = \frac{(\epsilon+p)}{p}\frac{\Delta p}{\Delta \epsilon},
    \label{eq:adiabaticindex}
\end{equation}
and  $H_0$ can be eliminated using: 
\begin{equation}
\begin{aligned}
 & \big[3M +\tfrac{1}{2}(l+2)(l-1)r+4 \pi r^3 p\big] H_0  = 8 \pi r^3 e^{-\nu} X \\ 
 & \;\;   - [\tfrac{1}{2}l(l+1)(M+4 \pi r^3 p) - \omega^2 r^3 e^{-2(\lambda + \nu)}] H_1  \\
 & \;\;   +  \big[\tfrac{1}{2}(l+2)(l-1)r-\omega^2 r^3 e^{-2\nu}   \\
 & \;\;   - r^{-1} e^{2\lambda} (M+4 \pi r^3 p)(3M - r + 4 \pi r^3 p)\big] K  .
\label{eq:relation_variables}
\end{aligned}
\end{equation}

The boundary conditions to be satisfied are: the perturbation functions must be finite everywhere, especially at $r=$0 where the system becomes singular; and the perturbed pressure must vanish at the surface of the star $r=R$ at any time, implying $\Delta p (r=R)$=0. We can write \citep{sotani2011}
\begin{equation}
    \Delta p = -r^l e^{-\nu} X,
    \label{eq:deltap_X}
\end{equation}
therefore $\Delta p (r=R)$=0 implies $X(r=R)=$0. For a given set of $l$ and $\omega$, there is only one solution which satisfies all of the boundary conditions.

To numerically solve the equations, we expand the solutions at $r=$0 and $r=R$ as suggested by Lindblom and Detweiler \citep{lindblomdetweiler1983,lindblomdetweiler1985} and corrected by Lü and Suen \citep{lusuen2011}. Concerning the surface of the star, we do not employ a polytropic atmosphere as in some previous works, i.e. we simply adopt $X=\mathrm{0}$ at the point where the pressure is effectively zero (we compared our calculations with previous works that considered polytropic atmospheres and we obtained the same results).
Notice that until now we have discussed only the boundary conditions for single-phase stars; for hybrid stars we will dedicate a special section.

In order to connect NS pulsations with GWs detected on terrestrial laboratories,  we need to know how the oscillation propagates until it reaches a distant observer.
In general, outside the star the perturbed metric describes a combination of outgoing and incoming GWs; however, we are particularly interested in purely outgoing radiation, representing the quasi-normal modes (QNMs) of the stellar model.

Outside the star the fluid quantities vanish and the perturbation equations reduce to the Zerilli equation \citep{zerilli1970,fackerell1971,chandrasekhardetweiler1975}
\begin{equation}
    \frac{d^2 Z}{dr^{*2}}+[\omega^2-V(r^*)]Z=0,
    \label{eq:Zerillieq}
\end{equation}
where the effective potential $V(r^*)$ is given by
\begin{eqnarray}
    V(r^*) &=& \frac{2(1-2M/r)}{r^3 (nr+3M)^2} \big[ n^2 (n+1)r^3+3n^2 M r^2  \nonumber \\
    && + 9nM^2 r + 9M^3 \big],
\end{eqnarray}
and $r^*$ is the ``tortoise'' coordinate, which can be written in terms of $r$ as
\begin{equation}
    r^* = r + 2M \log \left(\frac{r}{2M}-1 \right)
\end{equation}
with $n=(l-1)(l+2)/2$. 

In terms of $H_0(r)$ and $K(r)$, the Zerilli function $Z(r^*)$ and its derivative are
\begin{align}
    Z(r^*) &= \frac{k(r)K(r)-a(r)H_0(r)-b(r)K(r)}{k(r)g(r)-h(r)}, \label{eq:Zerilli_function} \\
    \frac{dZ(r^*)}{dr^*} &= \frac{h(r)K(r)-a(r)g(r)H_0(r)-b(r)g(r)K(r)}{h(r)-k(r)g(r)}, \label{eq:Zerilli_derivative}
\end{align}
where \citep{lusuen2011}
\begin{align}
    a(r) &= -(nr+3M)/[\omega^2 r^2 - (n+1)M/r], \\
    b(r) &= \frac{[nr(r-2M)-\omega^2 r^4 + M(r-3M)]}{(r-2M)[\omega^2 r^2 - (n+1)M/r]}, \\
    g(r) &= \frac{[n(n+1)r^2+3nMr+6M^2]}{r^2 (nr+3M)}, \\
    h(r) &= \frac{[-nr^2+3nMr+3M^2]}{(r-2M)(nr+3M)}, \\
    k(r) &= -r^2 / (r-2M).
\end{align}
The Zerilli equation has two linearly independent solutions $Z_+ (r^*)$ and $Z_- (r^*)$. They correspond to incoming and outgoing GWs respectively and the general solution for  $Z(r^*)$ is given by their linear combination
\begin{equation}
    Z(r^*) = A(\omega)Z_- (r^*) + B(\omega)Z_+ (r^*). \label{eq:Zer_expansion}
\end{equation}
At large radius, one can expand $Z_+$ and $Z_-$ as
\begin{align}
    Z_-(r^*) &= e^{-i \omega r^*} \sum_{j=0}^\infty \beta_j r^{-j}, \label{eq:Zer_minus_infty}\\
    Z_+(r^*) &= e^{i \omega r^*} \sum_{j=0}^\infty \overline{\beta}_j r^{-j}, \label{eq:Zer_plus_infty}
\end{align}
where $\overline{\beta}_j$ is the complex conjugate of $\beta_j$. Replacing Eq. \eqref{eq:Zer_minus_infty} (keeping terms to $j=\mathrm{2}$) into Eq. \eqref{eq:Zerillieq}, one obtains \citep{lusuen2011}
\begin{align}
    \beta_1 &= -i (n+1) \omega^{-1} \beta_0, \\
    \beta_2 = -\omega^2 & \left[ \frac{1}{2}n(n+1)-\frac{3}{2}iM \omega \left( 1+\frac{2}{n} \right) \right] \beta_0.
\end{align}

\section{Oscillating hybrid stars: junction conditions at the interface} \label{sec:junction}
 
In recent papers \cite{pereirafloreslugones2018,Pereira2}, we derived the junction conditions at the interface of a \textit{radially perturbed} hybrid star in the presence of slow and rapid phase conversions. 
In this section, we derive the junction conditions that hold at the sharp splitting surface of a hybrid star when the object is perturbed \textit{non-radially}.  We treat separately slow  (Sec. \ref{subsec:junctionslow}) and rapid  (Sec. \ref{subsec:junctionrapid}) phase conversions at the interface.

\subsection{Slow transitions} \label{subsec:junctionslow}

When a hybrid star is perturbed,  fluid elements in the neighbourhood of the sharp quark–hadron interface can be radially displaced and their pressures may become higher or lower than the phase transition pressure. However, a fluid element oscillating around the interface will not necessarily undergo a phase conversion. In fact, if the timescale of the process transforming one phase into another is much larger than the oscillation period (slow transitions), volume elements near the interface will simply comove with the splitting surface without changing their nature. In such a case there is no mass transfer across the interface. Since it is always possible to track down the elements near the splitting surface, then $\xi^r$ must be continuous across the interface, i.e. $[\xi^r]_-^+ = \xi^{r+}-\xi^{r-} = \mathrm{0}$. 
Since $\lambda$ and $r$ are continuous across the surface, we obtain from Eq. \eqref{eq:Xi} the first junction condition for slow transitions:
\begin{equation}
    [W]_-^+=0.
    \label{eq:Wjump_slow}
\end{equation}

Additionally, when a hybrid star oscillates,  the pressure on one side of the phase discontinuity keeps always the same value as on the other side, even if such value is different from the equilibrium one.  As a consequence, the Lagrangian change of the pressure must be continuous across the interface. Therefore, the second junction condition for slow transitions reads:  
\begin{equation}
    [\Delta p]_-^+ = 0.
    \label{eq:Deltapjump_slow}
\end{equation}

Notice that these junction conditions have already been used in several previous works (see e.g. \citep{sotani2001,sotani2011,floreslugones2014} and references therein) but without examining the role of the conversion speed at the interface. 
\subsection{Rapid transitions} \label{subsec:junctionrapid}
Rapid phase transitions happen when the characteristic timescale of the process transforming one phase into the other is much smaller than the timescale of the perturbations. As a limiting case we consider that a volume element near the phase-splitting boundary $\Sigma$ is converted instantaneously  from one phase to another when, due to perturbations, its pressure changes alternatively below and above the transition pressure $p_t$.
Since conversion rates are very fast, the pressure at the surface $\Sigma$  is always the same as for the unperturbed configuration, i.e.  $[p]_-^+= \mathrm{0}$ and, therefore,
\begin{equation}
    [\Delta p]_-^+= \mathrm{0}
    \label{eq:Deltapjump_fast}
\end{equation}
across the interface $\Sigma$. 

As done in \citep{pereirafloreslugones2018}, we will use only physical considerations to deduce the appropriate boundary condition for $\xi^r$ at $\Sigma$. We will demand that $\Sigma$ is well-localized, i.e., $[r_\Sigma]_-^+= \mathrm{0}$, where the $r_\Sigma^{\pm}$ is the radial position of the phase-splitting surface with respect to the radial coordinates above and below it, respectively. 
In equilibrium $\Sigma$ is at the position $r_\Sigma^{\pm}=R_0$. In the perturbed configuration, we should generically have $r_\Sigma^{\pm}=R_0 + \mathcal{A}(r_\Sigma^{\pm},\theta,\phi,t)$, where $\mathcal{A}^{\pm} \equiv \mathcal{A}(r_\Sigma^{\pm},\theta,\phi,t)$ are unknowns and of the order of $\xi^r$. 

Furthermore, the transition pressure is the equilibrium one, so at $r=r_\Sigma^{\pm}$ we have:
\begin{equation}
p(r_\Sigma^{\pm},\theta,\phi,t)=p_0 (R_0). 
\label{eq:physical_condition}
\end{equation}
On the lefthand side of Eq. \eqref{eq:physical_condition} we can use the definition of the Lagrangian displacement of the pressure $p(r,\theta,\phi,t)=p_0(r)+\Delta p (r,\theta,\phi,t) - \xi^r p_0'$, where $p_0 (r)$ stands for the pressure at $r$ in the unperturbed configuration. Thus, we can write:
\begin{equation}
\begin{aligned} 
    p(r_\Sigma^{\pm},\theta,\phi,t) = &  \; p_0(r_\Sigma^{\pm})+\Delta p (r_\Sigma^{\pm},\theta,\phi,t)  \\
           &  - \xi^r(r_\Sigma^{\pm},\theta,\phi,t) p_0'(r_\Sigma^{\pm}).
       \label{eq:P_Lagrangian}
\end{aligned}       
\end{equation}
Additionally, we can expand in series the quantity $p_0(r_\Sigma^{\pm})$ on the righthand side of the latter equation: 
\begin{equation}
 p_0 (r_\Sigma^{\pm}) \simeq p_0 (R_0) +  p_0^{\prime}(r_\Sigma^{\pm}) \mathcal{A}(r_\Sigma^{\pm},\theta,\phi,t). 
 \label{eq:series_expansion}
 \end{equation}
Replacing Eqs. \eqref{eq:P_Lagrangian} and \eqref{eq:series_expansion}  into Eq. \eqref{eq:physical_condition}, it follows that
\begin{equation}
\begin{aligned} 
& p_{0}(R_{0}) + p_0^{\prime}(r_\Sigma^{\pm})  \mathcal{A}(r_{\Sigma}^{\pm}, \theta, \phi, t)   \\
& + \Delta p(r_{\Sigma}^{\pm}, \theta, \phi, t)-\xi^{r}(r_{\Sigma}^{\pm}, \theta, \phi, t) p_{0}^{\prime}(r_{\Sigma}^{\pm})  = p_{0}(R_0) .
\end{aligned}
\end{equation}
In this equation, we eliminate $p_0 (R_0)$, we use Eq. \eqref{eq:Xi},  we write $\mathcal{A}$ and $\Delta p$ in terms of spherical harmonics, and we find: 
\begin{equation}
\begin{aligned} 
&    p_0^{\prime}(r_\Sigma^{\pm}) \mathcal{A} (r_\Sigma^{\pm}) Y_m^l (\theta,\phi) e^{i \omega t} + \Delta p(r_\Sigma^{\pm}) Y_m^l (\theta,\phi) e^{i \omega t} \\
&    - r_{\Sigma}^{l-1} e^{-\lambda} W(r_\Sigma^{\pm})Y^l_m(\theta,\phi) e^{i\omega t} p_0^{\prime}(r_\Sigma^{\pm}) = 0 .
\end{aligned}    
\end{equation}
Simplifying, we obtain 
\begin{equation}
   \mathcal{A} (r_\Sigma^{\pm})   = 
   - \frac{\Delta p(r_\Sigma^{\pm})}{ p_0^{\prime}(r_\Sigma^{\pm})}  + r_{\Sigma}^{l-1} e^{-\lambda} W(r_\Sigma^{\pm}).
  \label{eq:cal_A} 
\end{equation}
Now, from the condition $[r_\Sigma]_-^+ \equiv r_\Sigma^{+} - r_\Sigma^{-}  = \mathrm{0}$, we have $[\mathcal{A}(r)]_-^+ \equiv \mathcal{A}(r_\Sigma^{+})  - \mathcal{A}(r_\Sigma^{-})  = \mathrm{0}$, and then Eq. \eqref{eq:cal_A} reads:
\begin{equation}
    [W(r)]_-^+ = r_\Sigma^{-l+1} e^\lambda \left[\frac{\Delta p (r)}{p_0'}\right]_-^+ .
    \label{eq:Wjump_fast}
\end{equation}
Equations \eqref{eq:Wjump_fast} and \eqref{eq:Deltapjump_fast} are the junction conditions at the quark-hadron interface for rapid phase transitions. Notice the similarity of the radial case \citep{pereirafloreslugones2018} with the non-radial one. This occurs because of the freedom in writing the angular dependence through spherical harmonics.

Another form of the latter junction condition can be found replacing Eqs. \eqref{eq:deltap_X} and \eqref{eq:definitionX} into Eq. \eqref{eq:Wjump_fast} and taking into account that $H_0$ is continuous through the interface in view of the metric continuity:
\begin{equation}
    [V(r)]_-^+ = 0.
    \label{eq:jumpV_fast}
\end{equation}

\section{Nonexistence of discontinuity gravity modes in hybrid stars with rapid phase conversions} \label{sec:none}

Gravity modes are a consequence of buoyancy in a gravitational field and are intrinsically related with convective instabilities in stars. When a fluid element undergoes a small radial displacement outward, the star's gravity provides a force to restore the displaced element to its original location if  the displaced element's density is greater than that of the unperturbed fluid in the surroundings. When the displaced fluid element is of equal or lower density than the unperturbed fluid, gravity provides either no force (marginal stability) or a force to increase the displacement (instability to convection) \cite{finn1987}. A similar analysis is valid for a fluid element undergoing a small radial displacement downward.

To first order in small quantities, the relativistic buoyancy force per unit volume acting on a fluid
element displaced a small radial distance $\delta r$ is
\begin{equation}
f \equiv g(\epsilon+p) A e^\lambda \delta r 
\label{eq:f}
\end{equation}
where $-g$ is the gravitational acceleration in the radial direction measured by a stationary observer at $r$, $\gamma_0=\frac{(\epsilon+p)}{p}\frac{dp}{d\epsilon}$ is the adiabatic index in the unperturbed configuration  and $\gamma$ is given in Eq. \eqref{eq:adiabaticindex}. The quantity $A$ is the relativistic convective stability discriminant defined by
\begin{equation}
A \equiv e^{-\lambda} \frac{d p}{d r} \frac{1}{p} \left(\frac{1}{\gamma_{0}}-\frac{1}{\gamma}\right) .
\label{eq:A} 
\end{equation}
At the stellar layers where $A<0$, the local buoyancy force is restoring and the star is stable against convection. Conversely, the star is neutrally stable where $A=0$ and unstable where $A > 0$.
The buoyancy force density $f$ causes local fluid oscillations that are characterised by the  relativistic Brunt-Väisälä frequency $N^2 \equiv -Ag$,  which specifies the locally measured frequency with which a fluid element oscillates around its equilibrium position.  Different stability regimes can be identified by looking at the sign of $N^2$ \cite{finn1987}.

A zero temperature hybrid star  is chemically homogeneous everywhere except at the quark-hadron interface where the density changes discontinuously. Thus, a buoyancy force (and a $g$-mode) can only be expected at the interface, where we could have $\gamma \neq \gamma_{0}$. In fact, this is the case when conversions at the interface are \textit{slow}: the adiabatic index $\gamma_0$ governing the pressure-density relation is zero at the discontinuity, the adiabatic index $\gamma$  governing the perturbations remains finite there,  $A$ is different from zero, and a $g$-mode arises.
However, in the case of \textit{rapid} conversions, a displaced fluid element adjusts almost immediately its composition to its surroundings when it is pushed to the other side of the discontinuity. In this case, $\gamma_0$ is still zero at the interface, and so is $\gamma$ since complete thermodynamic equilibrium is maintained at all times with the unperturbed fluid. 
Therefore, we have $\gamma=\gamma_0=0$, which implies $A=0$, and there is no restoring force. 
In other words, since the displaced fluid element adjusts immediately its thermodynamic state to its surroundings, its density will be always the same as in the unperturbed fluid and gravity cannot provide a restoring force. 
As a consequence, the discontinuity $g$-mode has zero frequency for \textit{rapid} transitions.

\section{Discontinuity gravity modes in the case of slow phase conversions} \label{sec:results}

In this section, we study the $g$-mode arising from a quark-hadron phase discontinuity when phase conversions are slow. We first describe the EOSs adopted for both phases and then present the results of our calculations.

\subsection{Equations of State} \label{sec:eos}

\subsubsection{Hadronic Matter}

For hadronic matter we use an EOS based on nuclear interactions derived from chiral effective field theory (EFT), combined with constrains arising from the recent observation of high mass pulsars. In recent years, the development of chiral EFT has provided the framework for a systematic expansion for nuclear forces at low momenta allowing to constrain the properties of neutron-rich matter up to nuclear saturation density to a high degree. However, our knowledge of the EOS at densities greater than one to two times the saturation density is still insufficient due to limitations on both laboratory experiments and theoretical methods. Fortunately, the recent detection of very massive pulsars \cite{demorest2010,arzoumanian2018,antoniadis2013} with $\approx \mathrm{2}M_\odot$ puts stringent constraints on the nuclear EOS at supranuclear densities. Moreover, with the advent of GW observations of binary neutron star mergers \cite{abbott2017b,abbott2017c,abbott2017d}  additional constrains are emerging \citep{Raithel,Annala,Malik,Most,Fattoyev,Paschalidis:2017qmb,Tews:2018iwm,Christian:2019qer}. 

The EOS at subnuclear densities can be extended in a general way to higher densities using piecewise polytropic EOSs and requiring non-violation of causality and consistence with the observation of $2 M_\odot$ pulsars. In Ref. \cite{hebelerlattimer2013}, hadronic matter at densities above $\rho_1 = 1.1 \rho_{0}$ ($\rho_0 =\mathrm{2.7\times10^{14} g/cm^3}$) is described by a set of three polytropes which are valid, respectively, in three consecutive density regions. This general polytropic extension leads to a very large number of EOSs, which verify the physical and observational constraints mentioned above. For use in astrophysical simulations,  Ref. \cite{hebelerlattimer2013}  provides detailed numerical tables for three representative EOS labeled as \textit{soft}, \textit{intermediate} and \textit{stiff}. In order to ensure that our hybrid configurations verify the $2 M_\odot$ constrain we adopt only the \textit{intermediate} and \textit{stiff} parametrizations of Ref. \cite{hebelerlattimer2013}. Below $\rho_{\textrm{crust }}=\rho_{0} / 2$ we use the Baym, Pethick and Sutherland EOS with the extension to $\rho=\mathrm{5 \times 10^{14} \ g/cm^3}$ of Baym, Bethe and Pethick \cite{BPS1971}. For more details, see Ref. \cite{hebelerlattimer2013}.

\subsubsection{Quark Matter}

For quark matter we consider a generic MIT bag model which is defined by the following grand thermodynamic potential \cite{alford2005}: 
\begin{equation}
\Omega = -\frac{3}{4 \pi^2} a_4 \mu^4 +  \frac{3}{4 \pi^2} a_2 \mu^2 + B,
\label{eq:eosMIT}
\end{equation}
where $\mu = (\mu_u+\mu_d+\mu_s)/3$ is the quark chemical potential and  $a_{4}$, $a_{2}$,  and  $B$ are  free parameters independent of $\mu$.  Since the quark matter EOS is used here essentially in the high density regime, we have  neglected the electron contribution (see discussion in Ref. \cite{pereirafloreslugones2018}).

The above phenomenological model is interesting because it allows exploring several aspects of dense quark matter. 
The influence of strong interactions on the pressure of the free-quark Fermi sea is roughly taken into account by the parameter $a_4$, where  $0\leq a_4 \leq1$, and $a_4=1$ indicates no correction to the ideal gas \cite{alford2005}. 
The standard MIT bag model is obtained for $a_4=\mathrm{1}$ and $a_2=m_s^2$, being $m_s$ the mass of the strange quark. 
The effect of the color superconductivity phenomenon in the Color Flavor Locked (CFL) phase can be explored setting $a_2=m_s^2-4\Delta^2$, being $\Delta$ the energy gap associated with quark pairing \cite{alford2005,pereirafloreslugones2018}. The bag constant $B$ is related to the confinement of quarks, representing in a phenomenological way the vacuum energy \cite{vasquez2010}.

From Eq. \eqref{eq:eosMIT}, we can obtain all thermodynamic quantities, such as the pressure $p=-\Omega$, the baryon number density:
\begin{equation}
n_B=-\frac{1}{3} \frac{\partial \Omega}{\partial \mu}=\frac{1}{\pi^2} a_4 \mu^3 - \frac{1}{2\pi^2} a_2 \mu,
\label{eq:nb_MIT}
\end{equation}
and the energy density
\begin{equation}
\epsilon = \Omega + 3 \mu n_B   = \frac{9}{4 \pi^2} a_4 \mu^4 - \frac{3}{4 \pi^2} a_2 \mu^2 + B.
\label{eq:rho_mu_MIT}
\end{equation}
The chemical potential can be written as a function of pressure, 
\begin{equation}
\mu^2 = \frac{1}{2} \left[\frac{a_2 + a_2 \sqrt{1+\frac{16 \pi^2 a_4}{a_2^2}(\epsilon-B)} }{3 a_4}\right] ,
\label{eq:muquad_rho_MIT}
\end{equation}
which allows finding the EOS $p=p(\epsilon)$:
\begin{equation}
p = \frac{\epsilon-4B}{3}  - \frac{a_2^2}{12 \pi^2 a_4} \left[1+ \sqrt{1+\frac{16 \pi^2 a_4}{a_2^2} (\epsilon-B)} \right].
\label{eq:p_rho_MIT}
\end{equation}
Depending on the values of $a_2$, $a_4$ and $B$, either hybrid stars or strange stars may be described by this model. For more details see Ref. \cite{pereirafloreslugones2018} and references therein.

\begin{table}[tb]
\centering
\caption{Combinations of EOS' parameters adopted to construct hybrid stars models. }
\bigskip 
\label{table:hybridmodels}
\begin{tabular}{c||c|ccc}
\hline \hline
\textbf{Hybrid } & \textbf{Hadronic EOS} &   \multicolumn{3}{c}{\textbf{Quark EOS}}   \\ 
     \textbf{model}        &       & $B$[MeV/fm$^3$] & $a_2^{1/2}$[MeV] & $a_4$\\ 
\hline \hline 
 Hyb-S1 & Stiff  &92.55 &150 &0.7 \\
 Hyb-S2 & Stiff   &29.28 &150 &0.5 \\
 Hyb-S3 & Stiff   &74.28 &150 &0.5 \\
 Hyb-S4 & Stiff   &46.16 &100 &0.5 \\
 Hyb-I1 & Intermediate   &92.55 &150 &0.7 \\
 Hyb-I2 & Intermediate   &41.16 &100 &0.5 \\ \hline
\end{tabular}
\end{table}

\subsubsection{Hybrid Matter}

In order to describe hybrid stars, we combine the hadronic and the quark EOSs described above. As mentioned before, we assume that matter has a first order quark-hadron phase transition with a sharp density discontinuity at the pressure $p_t$. 
Once the model parameters are chosen, the transition pressure $p_t$ is found by requiring that the  Gibbs free energy per baryon $g$ of both phases is the same at $p_t$:  
\begin{equation}
g_H(p_t) = g_Q(p_t),
\label{eq:Gibbs_condition}
\end{equation}
where $g = ( \sum_i \mu_i n_i ) / n_B$,  being $\mu_i$  the chemical potential of particle species $i$, $n_i$ their number density, and  $n_B= \tfrac{1}{3} \sum_i n_i$ the baryon number density of each phase.
The quark phase is energetically preferred for $p>p_t$ and the hadronic phase for $p<p_t$. We have chosen the EOS parameters in order to allow the existence of hybrid stars with $M > 2 M_{\odot}$.   The choice of parameters employed in the present paper is presented in Table \ref{table:hybridmodels}.

\subsection{Results}
\label{sec:results_slow}

\begin{figure}[tb]
    \centering
    \includegraphics[width=0.45\columnwidth]{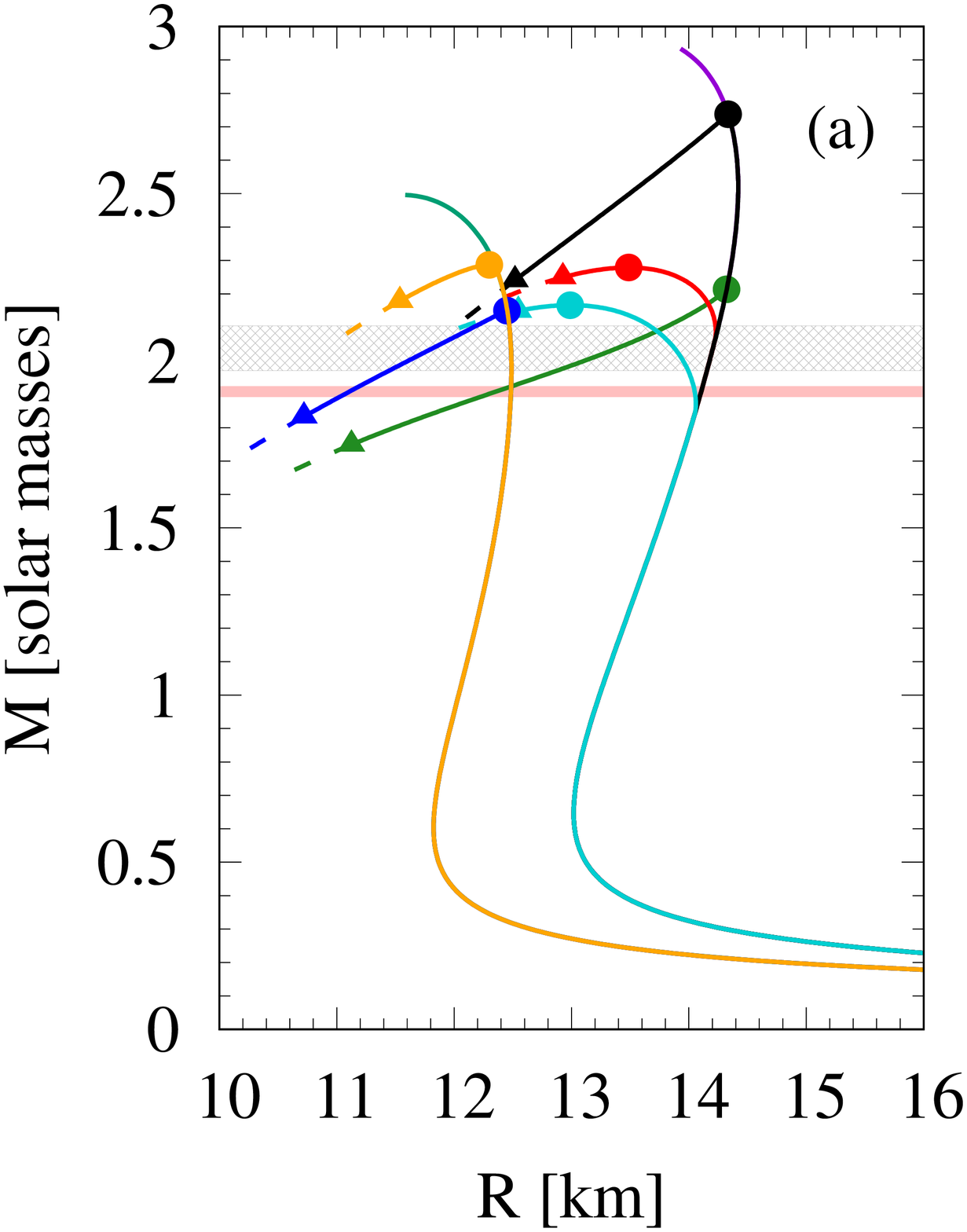}
    \includegraphics[width=0.45\columnwidth]{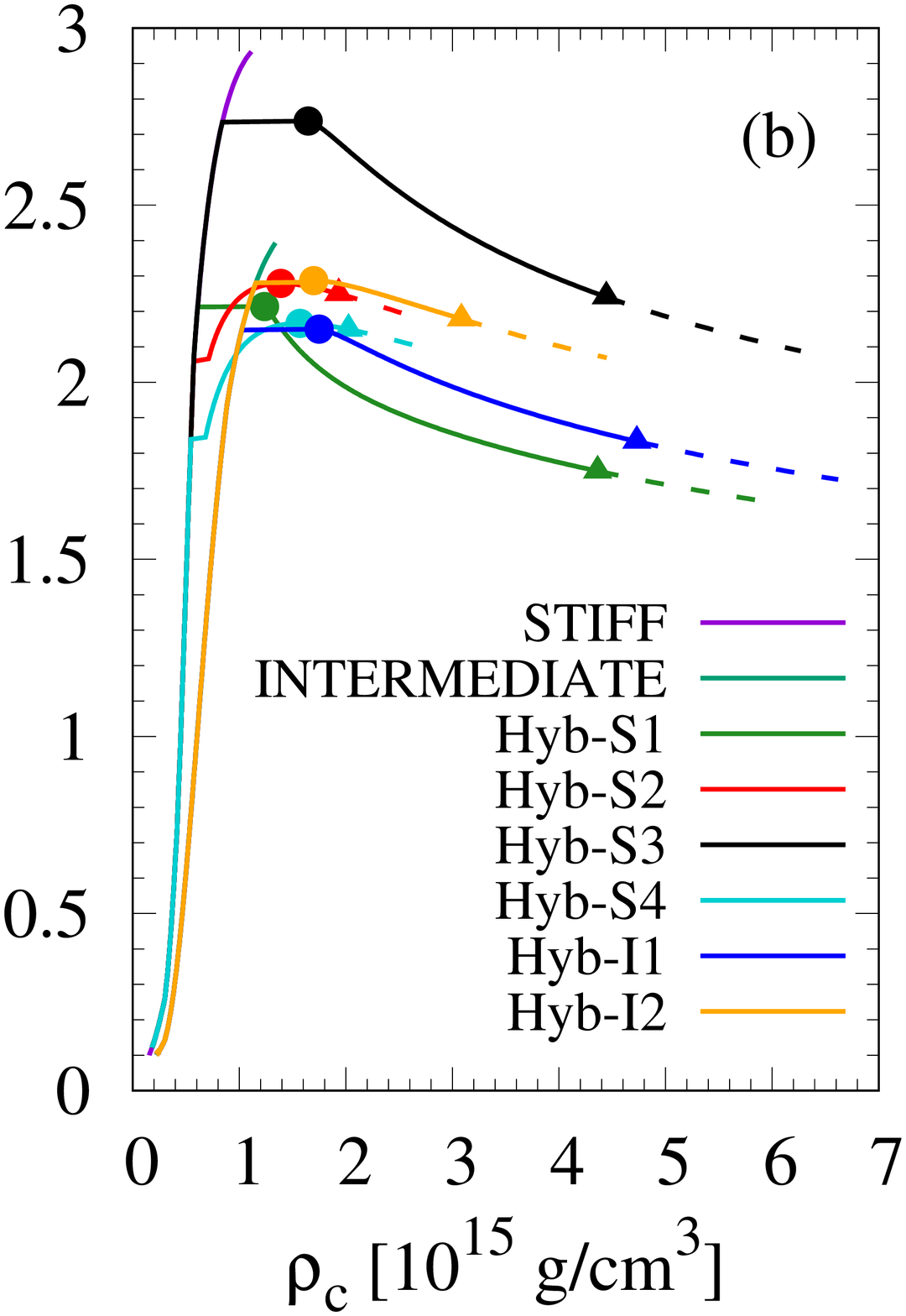}        
    \caption{Mass--radius relationship (a) and stellar mass as a function of the central density (b) for the hybrid models of Table \ref{table:hybridmodels}. In both panels, round dots indicate the maximum mass and triangular dots mark the last stable hybrid configuration for which the frequency of the fundamental radial oscillation mode vanishes in the case of slow phase conversions. Extended stable branches begin at round dots and end at triangular dots (only for slow conversions). The upper horizontal band on panel (a) corresponds to the observed mass of the pulsar PSR J0348+0432 and the lower horizontal band to PSR J1614-2230 \cite{arzoumanian2018,antoniadis2013}. } 
    \label{fig:staticconfig}
\end{figure}


We have shown in previous works  \cite{pereirafloreslugones2018} that, in the case of slow phase conversions, \textit{stable} hybrid configurations could exist even in some cases for which $\partial M/ \partial \epsilon_c <0$.  This introduces an \textit{extended family} of stable stars that begins at the maximum mass configuration and extends up to the  \textit{terminal configuration} at which the frequency of the fundamental radial oscillation mode vanishes.  This new \textcolor{blue}{family} can be seen in Fig. \ref{fig:staticconfig}. The round dots indicate the maximum mass stars while the triangular ones indicate the terminal configuration for each model. 

Notice that all the hybrid models of Table \ref{table:hybridmodels} allow the existence of twin objects, i.e. couples of stars with the same gravitational mass but different radii. This is a very relevant signature that may be used to scrutinise the internal composition of compact objects \cite{Paschalidis:2017qmb,Christian:2019qer,burgiodrago2018,alvarezcastillo2019,montanatolos2019,lisedrakianalford2020}. In fact, new missions probing neutron star radii such as NICER will be able to measure NS radii with 5\%-10\% of uncertainty while the future eXTP is expected to have even better precision. As an example, let us consider the Hyb-S2 model at a mass of $\mathrm{2.25} M_\odot$: the difference in radius between the hybrid star in the standard branch and the one with the same mass in the extended branch is around 8\%. 
For the Hyb-I1 model at $\mathrm{2} M_\odot$, such difference is  $\sim 9 \%$. Moreover, for the Hyb-S1 model, the difference in radius between the terminal configuration ($\mathrm{1.75}M_\odot$) and its hadronic twin is 26\%. These examples show that the extended branch can be observationally constrained by systematic mass and radii measurements of forthcoming missions.

\begin{figure}[tb]
    \centering
     \includegraphics[width=0.50\columnwidth,angle=-90]{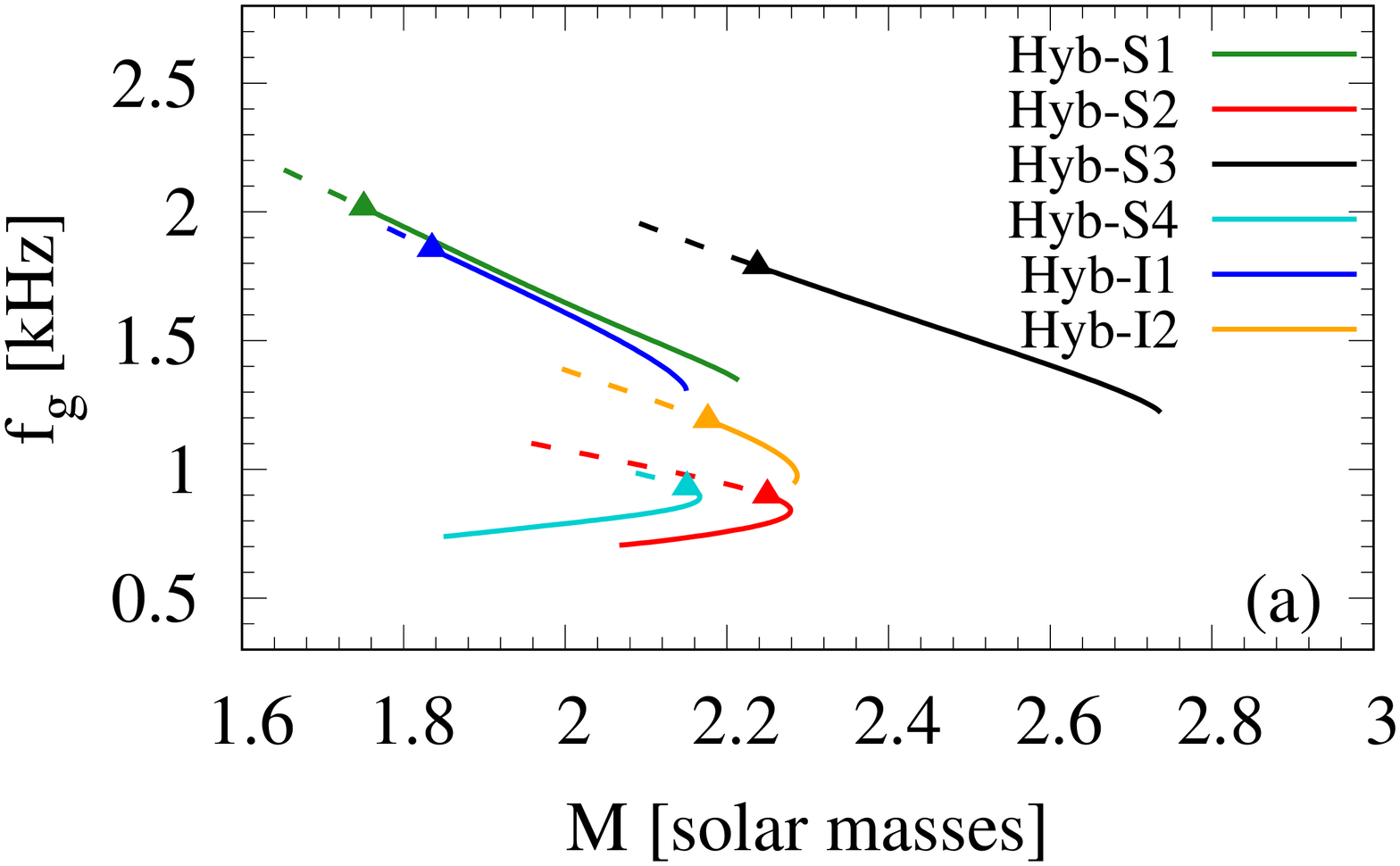}
      \includegraphics[width=0.50\columnwidth,angle=-90]{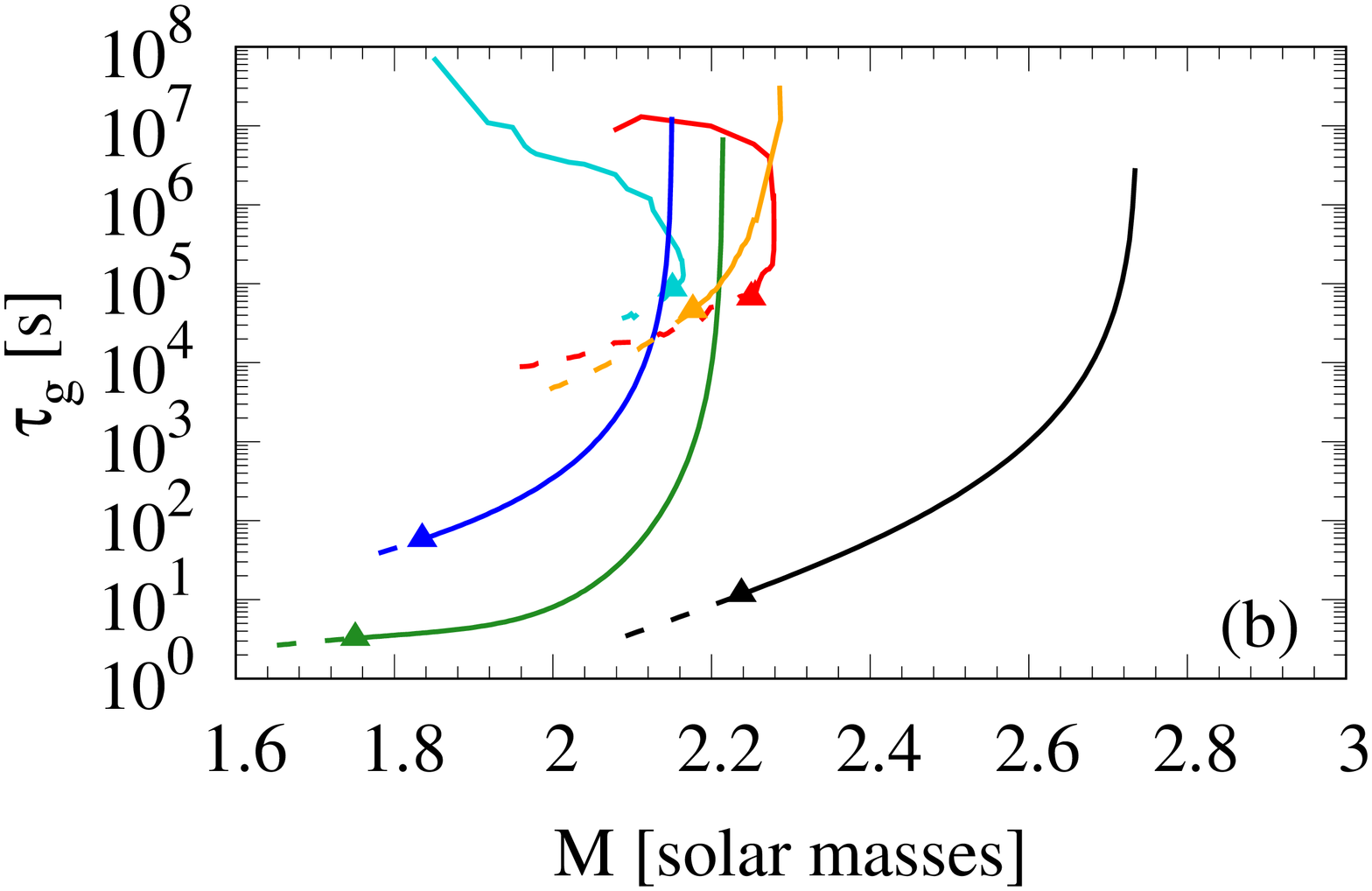}  
   \caption{Frequency (a) and damping time (b) of the $g$-mode for the hybrid configurations presented in Fig.  \ref{fig:staticconfig}. The triangular dots indicate the terminal configuration in the case of slow conversions. } 
    \label{fig:gmode}
\end{figure}

In Fig. \ref{fig:gmode} we show our results for the pulsation frequency ($f=\mathrm{Re}\{\omega\}/2\pi$) and the damping time ($\tau=\mathrm{1/Im}\{\omega\}$) of $g$-modes for the stable hybrid stellar models presented in Fig. \ref{fig:staticconfig}.  For discontinuity $g$-modes of the standard branch (that is, with $\partial M/ \partial \epsilon_c > 0$) we find values that are in agreement with previous results in the literature \cite{miniutti2003,sotani2011}, i.e. frequencies in the range $0.5-1 \, \mathrm{kHz}$ and very long damping times (see the lower branch of models Hybrid--S2, Hybrid--S4 and Hybrid--I2 in Fig. \ref{fig:gmode}a).  However, for $g$-modes of the extended branch (i.e. with  $\partial M/ \partial \epsilon_c < 0$) we find significantly larger oscillation frequencies and much shorter damping times.  In fact, for models Hybrid--S2, Hybrid--S4 and Hybrid--I2, $g$-modes of the extended branch are above the ones  of the standard branch and take values around 1kHz. For models Hybrid--S1, Hybrid--S3 and Hybrid--I1, all hybrid models belong to the extended branch and their $g$-mode frequencies are in the range $1.2-2 \, \mathrm{kHz}$  (see Fig. \ref{fig:gmode}a) while damping times can be as short as some seconds (see Fig. \ref{fig:gmode}b). 

\begin{figure}[tb]
\centering
\vspace{1.2cm}
\includegraphics[width=0.75\columnwidth,angle=0]{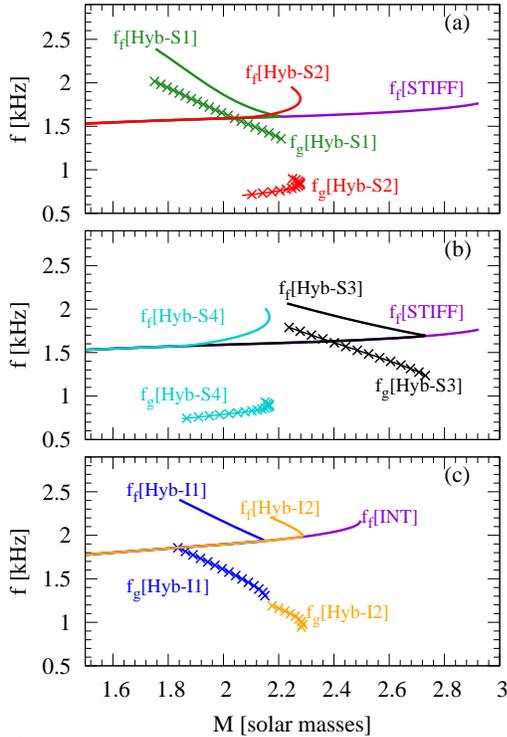}
\caption{Frequency of $f$- and $g$-modes for the models considered in this work. The asterisk-lines indicate the frequency of $g$-modes, while solid lines represent $f$-modes. The curves for the $f$-mode include results for both, hadronic and hybrid stars.}
\label{fig:compfg_freq}
\end{figure}

\begin{figure}[tb]
\centering
\vspace{1cm}
\includegraphics[width=0.75\columnwidth,angle=0]{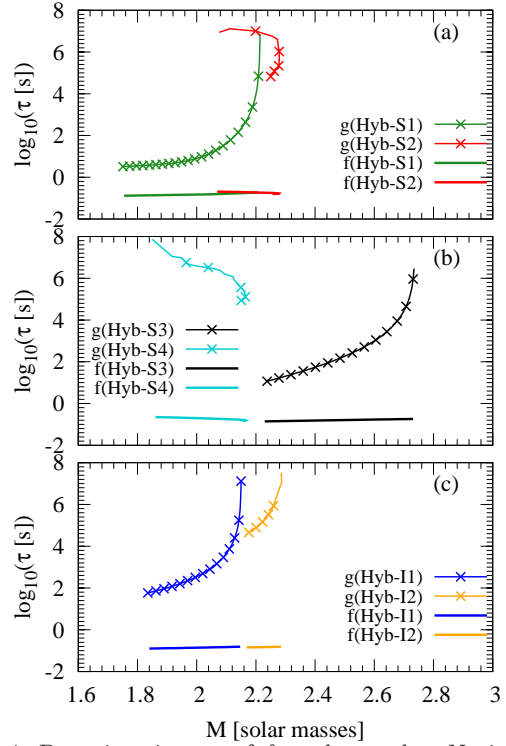}
\caption{Damping times $\tau$ of $f$- and $g$-modes. Notice that $\tau_{\mathrm{f}}$ and $\tau_{\mathrm{g}}$ differ by several orders of magnitude.}
\label{fig:compfg_damp}
\end{figure}

Since frequencies around $\mathrm{2 \, kHz}$ are typical of the fundamental mode of NSs, it is important to compare systematically the frequencies of both $f$- and $g$-modes of our hybrid configurations.  As seen in  Fig. \ref{fig:compfg_freq}, within each model $f_{\mathrm{f}}$ is always larger than $f_{\mathrm{g}}$ of a NS with the same gravitational mass (as it must be). But for some models, e.g. Hyb-S1, the difference between $f_\mathrm{f}$ and $f_\mathrm{g}$ is small, which may make difficult their observational discrimination. However, since $\tau_{\mathrm{g}}$ is several orders of magnitude larger than  $\tau_{\mathrm{f}}$ (see Fig. \ref{fig:compfg_damp}), both modes would be clearly differentiated if damping times were observed.

\begin{figure}[tb]
    \centering
     \includegraphics[width=0.50\columnwidth,angle=-90]{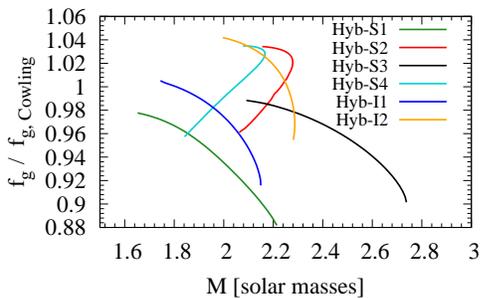} 
    \caption{Comparison between the Cowling approximation and the full formalism for frequencies of the $g$-mode.}
    \label{fig:Cowling}
\end{figure}

A brief comment on some numerical issues is in order. Our calculations have been done using the standard algorithm of Lindblom and Detweiler \cite{lindblomdetweiler1983,lindblomdetweiler1985}. As already emphasised by Finn \cite{finn1986,finn1987,finn1988},  both the effort and the error involved in the integration of $g$-modes may become large with such method. In fact, since in many cases the imaginary part of the eigenfrequency is fractionally too small compared to the real part,  a small fractional error in the real frequency can seriously affect the estimate of the damping time.  To circumvent this difficulty, we have first calculated the frequency of $g$-modes using the Cowling approximation \cite{cowling1941,mcdermott1983} and have used these results as initial values for the full calculation.  With such approach, we were able to determine $f_{\mathrm{g}}$ with high precision but in some cases it was difficult to resolve numerically the value of $\tau_{\mathrm{g}}$ with arbitrary precision. As a consequence, some of the curves shown in Figs. \ref{fig:gmode}b and \ref{fig:compfg_damp} are not smooth. 
As a byproduct of our calculations, we present some results obtained within the Cowling approximation that allow assessing its accuracy. 
The Cowling approximation, first developed for the study of Newtonian stars \cite{cowling1941} and subsequently adapted for the investigation of relativistic stars \cite{mcdermott1983}, arises when one neglects all metric perturbations in the full equations of Sec. \ref{sec:NRO} and it strongly simplifies the calculation of the frequency of quasi-normal modes  \cite{floreslugones2014}.  In Fig. \ref{fig:Cowling}  we show the ratio between the frequency $f_{\mathrm{g}}$ calculated within the full formalism and the frequency $f_{\mathrm{g, Cowling}}$ obtained within the Cowling approximation. For lower masses the approximation tends to be reasonably good  but for larger ones the difference can be as large as  $\sim 10\%$.

Now, let us focus on the detectability of the modes calculated in this work. It is possible to estimate the minimum energy that must be released through a mode in order to be detected by a given GW observatory according to the formula \cite{anderssonferrari2011,kokkotasapostolatos2001}
\begin{equation}
\begin{aligned}
    \frac{E_\mathrm{GW}}{M_\odot c^2} = & \;  3.47 \times 10^{36} \left(\frac{S}{N} \right)^2 \frac{1+4Q^2}{4Q^2}  \times \\
 &  \times  \left(\frac{D}{10 \mathrm{kpc}} \right)^2  \left(\frac{f}{1 \mathrm{kHz}} \right)^2 \left( \frac{S_n}{\mathrm{1 Hz^{-1}}} \right),
\end{aligned}    
    \label{eq:energyGW}
\end{equation}
where $E_\mathrm{GW}$ is the energy emitted in the form of GWs, $S/N$ is the signal-to-noise ratio, $Q=\pi f \tau $ is the quality factor, $D$ the distance to the source, $f$ the frequency, $\tau$ the damping time and $S_n$ the noise power spectral density of the detector. 

We consider a detector with $S_n^{1/2} \sim 2 \times 10^{-23} \, \mathrm{Hz}^{-1/2}$ which is representative of the Advanced LIGO-Virgo at $\sim$kHz \cite{abbott2017b}, and  another one with $S_n^{1/2} \sim 10^{-24} \, \mathrm{Hz}^{-1/2}$ which is illustrative of the planned third-generation ground-based Einstein Observatory at the same frequencies \cite{abbott2017future}. Taking $S/N=8$ we calculated the minimum energy $E_\mathrm{GW}$ that a NS must release through a mode in order to be detected at a  distance $D \sim10$ kpc (NS in our Galaxy) and $D \sim 15$ Mpc (NS at the Virgo cluster).

\begin{figure}[tb]
\vspace{0.5cm}
\includegraphics[width=0.45\columnwidth]{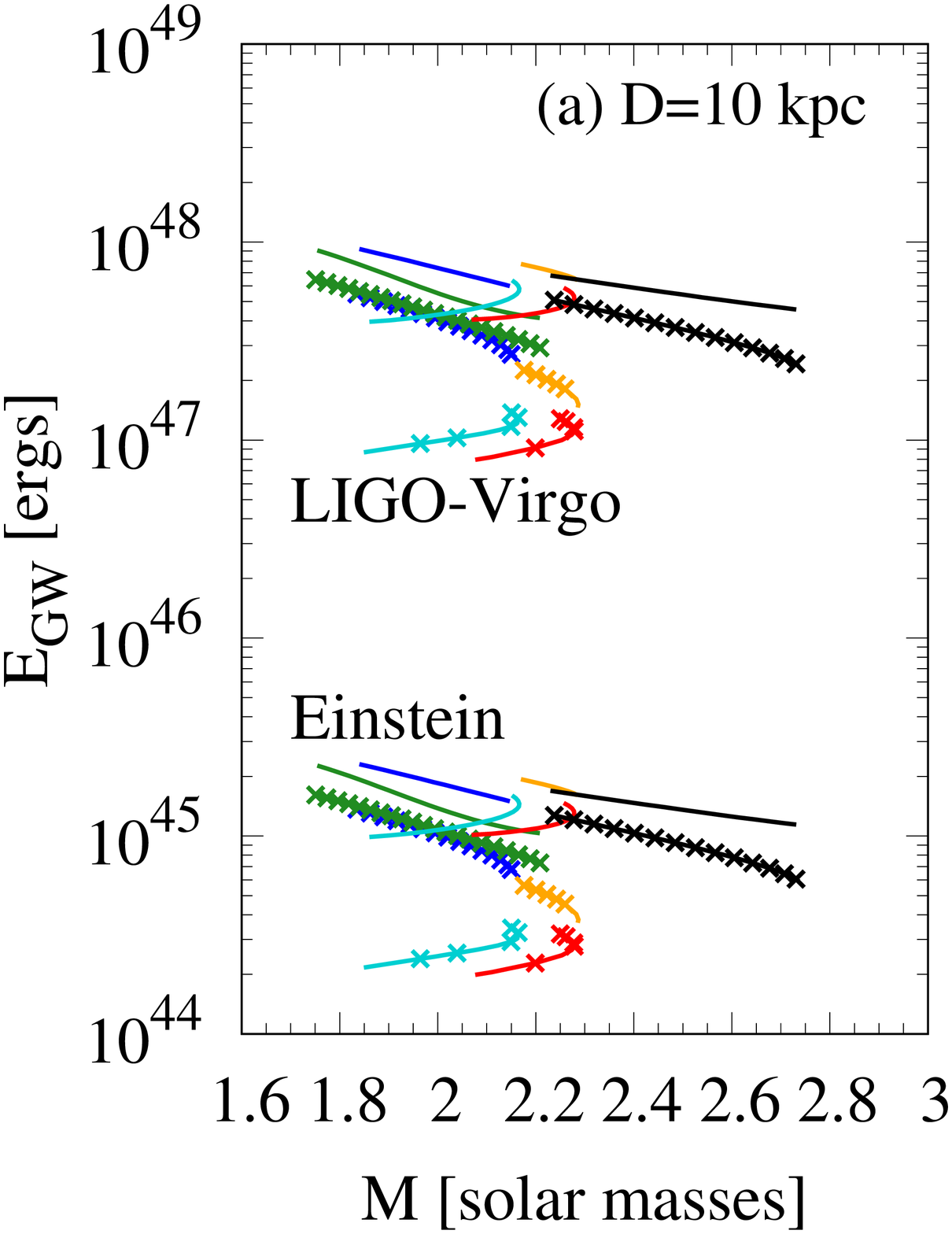}
\includegraphics[width=0.45\columnwidth]{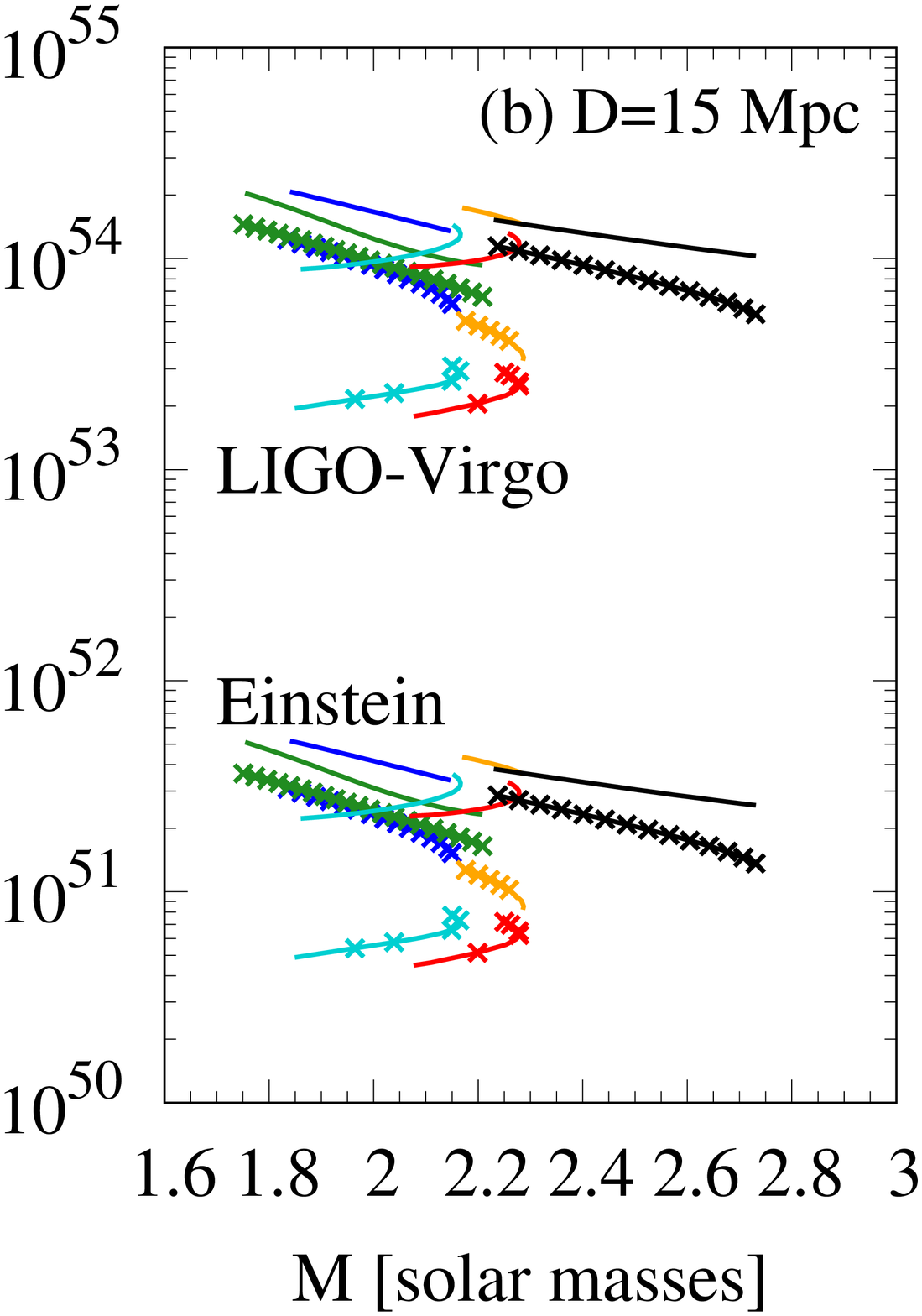}        
\caption{Minimum energy that must be released by a NS through GWs of the $f$ and $g$-modes in order to be detected by the Advanced LIGO/Virgo and the Einstein Observatory at a distance of (a) 10 kpc and (b) 15 Mpc.  The labels are the same as in Figs. \ref{fig:compfg_freq} and \ref{fig:compfg_damp}. }
\label{fig:energyGW}
\end{figure}

Our results are shown in Fig. \ref{fig:energyGW} and show that $E_\mathrm{GW}$ for $g$-modes is lower than for $f$-modes. However, in order to assess the relevance of each mode in GW emission, one must analyse several factors, being the amount of energy that can be stored in a given mode the most important. Furthermore, the amount of energy that can be channeled through GWs, depends on other dissipative processes that take energy away from the star, e.g. neutrino diffusion and viscosity (for the case of a newly born, hot star). Numerical simulations of extremely energetic processes, like core collapse to a NS or binary coalescence leading to NS formation, indicate that the $f$-mode is the most excited \cite{ferrarigualtieri2008}. However, further work should be done regarding these astrophysical simulations in view of the possible existence of the new extended stable branch discussed in this work. Since $g$-modes of this new branch have a significantly larger frequency, one may wonder whether they could carry more energy than $g$-modes of the standard branch. 

Even so, the results presented in Fig. \ref{fig:energyGW} look promising. Following a catastrophic astrophysical event such as a supernova collapse, a binary coalescence or a conversion of a hadronic star into a hybrid star, one expects that a strongly pulsating compact star will be created (if the event doesn't end with the formation of a black hole). Although it is yet uncertain how much energy will be radiated through the oscillation modes, one can reasonably expect that the energy stored in stellar pulsations is some fraction of the kinetic energy of the formation event. In the case of a typical core collapse supernova, the total released energy is $\sim10^{53}$ ergs while the kinetic energy of mass ejecta is $\sim 10^{51}$ ergs.  Thus, the observation of  $g$-mode GWs from a Milky Way event looks feasible, since Fig. \ref{fig:energyGW}a shows that the minimum detectable energy is in the range $\sim 10^{47}$--$10^{48}$ ergs for Advanced LIGO-Virgo. The Einstein Telescope, with a threshold in the range $\sim 10^{44}$--$10^{45}$ ergs for Galactic $g$-mode GWs is much more encouraging.
Giant flares of Soft Gamma Repeaters (SGR) may be another detectable source of GWs. In the magnetar model, SGRs are highly magnetised NSs with surface magnetic fields around $10^{15}$ G. During giant flares, up to $\sim 10^{47}$ ergs may be released in $\gamma$-rays as a consequence of a strong rearrangement of the magnetic field probably  leading to crustal deformations and cracking with the potential excitation of non-radial pulsation modes. According to Fig. \ref{fig:energyGW}a a detection of a galactic SGR with Advanced LIGO-Virgo requires the energy released in  $g$-mode GWs to be of the same order of the one released in $\gamma$-rays. In the case of the Einstein Observatory, the minimum required energy is $100-1000$ times smaller.
For completeness, the curves for sources in the Virgo cluster of galaxies are shown in Fig. \ref{fig:energyGW}b.

\section{Summary and Conclusions} \label{sec:conclusion}

In this paper we investigated the role of slow and rapid phase conversions on non-radial quasi-normal modes of hybrid stars. To this end, we derived the junction conditions that hold at the sharp interface of a perturbed hybrid star in the case of slow conversions (Eqs. \eqref{eq:Wjump_slow} and \eqref{eq:Deltapjump_slow}) and rapid conversions (Eqs. \eqref{eq:Deltapjump_fast} and \eqref{eq:Wjump_fast}).  

After that, we focused on the discontinuity $g$-mode because of its relevance as a fingerprint of a sharp quark-hadron interface at the compact star interior. 

In Section \ref{sec:none} we analysed the physical mechanism that suppresses the existence of discontinuity $g$-modes when phase conversions at the interface are rapid. In this case, a displaced fluid element near the phase splitting surface adjusts almost immediately its composition to its surroundings when it is pushed to the other side of the discontinuity. Since it is always in equilibrium with its environment, its density will be always the same as in the unperturbed fluid and gravity cannot provide a restoring force. In fact, the relativistic buoyancy force per unit volume acting on a displaced fluid element (see Eq. \eqref{eq:f}) vanishes for rapid conversions because the adiabatic index $\gamma_0$ governing the pressure-density relation and the adiabatic index $\gamma$ governing the perturbations are both zero at the discontinuity. Therefore, the discontinuity $g$-mode has zero frequency if phase conversions are \textit{rapid}.

In the case of \textit{slow} conversions, a buoyancy force and a $g$-mode arise at the interface because the adiabatic index $\gamma$  governing the perturbations remains finite there. In Section \ref{sec:results_slow}, $g$-modes were analysed using the EOSs for hadronic and quark matter presented in Sec. \ref{sec:eos}. Concerning slow conversions, notice  that we have shown in previous works \cite{pereirafloreslugones2018,Mariani:2019vve} that a new branch of \textit{stable} hybrid configurations arises for which $\partial M/ \partial \epsilon_c <0$.  Such \textit{extended branch}  begins at the maximum mass configuration and extends up to the  \textit{terminal configuration} at which the frequency of the fundamental \textit{radial} oscillation mode vanishes. Our results show that $g$-modes of the standard branch (that is, with $\partial M/ \partial \epsilon_c > 0$) have frequencies and damping times in  agreement with previous results in the literature \cite{miniutti2003,sotani2011}, i.e. frequencies in the range $f \sim 0.5-1 \, \mathrm{kHz}$ and very long damping times. However, for $g$-modes of the extended branch we obtain significantly larger frequencies (in the range $1-2 \, \mathrm{kHz}$) and much shorter damping times (few seconds in some cases).  

Finally, we discussed the detectability of $g$-mode GWs with present and planned GW observatories. The minimum released energy in $g$-mode GWs for a source at a galactic distance (10 kpc) is in the range $\sim 10^{47}$--$10^{48}$ ergs for Advanced LIGO-Virgo and in the range $\sim 10^{44}$--$10^{45}$ ergs for the  Einstein Telescope. These results suggest that the detection of $g$-mode GWs from nearby core collapse supernova, compact star mergers and even SGRs is feasible, and that $g$-modes are a promising  tool for the search of sharp quark-hadron discontinuities at the deep interior of compact stars.

\acknowledgements
L. Tonetto acknowledges the  Funda\c c\~ao de Amparo \`a Pesquisa do Estado de S\~ao Paulo (FAPESP) under grant No. 2018/04281-8. G. Lugones acknowledges the Brazilian agency Conselho Nacional de Desenvolvimento Cient\'{\i}fico e Tecnol\'ogico (CNPq) for financial support. 
We thank Jonas Pedro Pereira for helpful discussions.

\bibliography{paper1.bib}
\bibliographystyle{apsrev4-1}
\end{document}